# Accurate humidity and pH synchronized measurement with temperature compensation based on polarization maintaining fiber

Jia Liu[1]*, Jiawen Zhang[3], Xiyu Liu[1], Qi Meng[1], Riming Xu[2], Jin Wang[2]*
1. College of Information Science and Engineering, Shandong Agriculture and Engineering University, Shandong, 250100, China
2. Physical Sciences and Engineering Division, King Abdullah University of Science and Technology, Thuwal 23955-6900, Saudi Arabia
3. School of Chemical Engineering and Light Industry, Guangdong University of Technology, Guangzhou 510006, Guangdong, China.

**Abstract**
Real-time and accurate monitoring of humidity and pH is of great significance in daily life and industrial production. Existing humidity and pH measurement suffer from limitations such as low sensitivity, signal crosstalk, complex system structures, and inability to achieve real-time monitoring. In this work, the surface of a polarization maintaining fiber (PMF) was functionalized with a composite humidity-sensitive polymer composed of polyvinyl alcohol (PVA) and carbon nanosheets (CNs). A humidity-sensitive film with a microporous structure was prepared on the PMF cladding through high-temperature rapid film formation and laser processing, enhancing humidity sensitivity and stability. To enable pH sensing, poly(allylamine hydrochloride) (PAH) and poly (acrylic acid) (PAA) were successively adsorbed onto the PMF surface via electrostatic self-assembly, forming a pH-sensitive nanofilm structure. By connecting a temperature-compensated PMF within the same Sagnac loop and combining it with a multi-wavelength matrix, simultaneous real-time monitoring of humidity, pH, and temperature was achieved, effectively solving the issue of temperature crosstalk and extending toward a universal optical fiber multi-parameter measurement platform.
**Key words:** multi-parameter measurement, cascade, polarization maintaining fiber, humidity, temperature, PH.

## 1. Introduction

Humidity and acidity are recognized as two fundamental environmental parameters that critically influence diverse domains, including industrial manufacturing, agricultural production, biomedical engineering, and environmental monitoring. Fluctuations in humidity markedly affect material properties, product durability, and process efficiency, whereas variations in acidity serve as indicators of chemical reactivity, corrosion propensity, and contamination levels. In particular, under complex or confined environmental conditions, humidity and acidity often exhibit coupled variations, jointly impacting system stability and the operational lifespan of equipment [1-3]. Developing a low-cost humidity and pH sensor with a faster and more accurate response assumes great importance. The measurement of electronic sensors is limited by the transmission and reception of signals, requiring manual on-site signal collection [4-5]. This cannot meet the real-time monitoring of signals in complex and dangerous environments. The measurement of optical fiber sensors is increasingly used for real time and remote monitoring in harsh environments due to the advantages of all-optical communication, quick response and low-loss [6-7]. The sensor and spectrometer can be placed at a long distance apart, without requiring manual on-site signal collection and subsequent data transmission. Combining with hydrogel, carbon nanotubes, polyvinyl alcohol and other humidity-sensitive materials, various kinds of optical fiber sensors have been reported for humidity sensing, such as long-period gratings [8], photonic crystal fibers [9], Fabry–Perot interferometer [10], fiber Bragg gratings [11]. Combining poly (allylamine hydrochloride) (PAH) and poly(acrylic acid) (PAA) multilayers, Poly DiallylDimethylAmmonium chloride/PAA thin films, Brilliant Yellow (BY) dye coatings, and various polymer composites [12-13], a range of optical fiber structures have been developed for pH sensing applications. Representative examples include pH sensors based on tilted fiber Bragg gratings [14], straight fiber sensors fabricated by layer-by-layer self-assembly [15].

However, optical fiber sensors are often limited by low sensitivity and poor stability. Therefore, the selection of an appropriate sensitive material and film preparation technology is a crucial aspect in achieving and enhancing sensitivity. Furthermore, the accuracy of humidity and pH measurement is significantly compromised by the cross sensitivity inherent in optical fiber, as it is susceptible to other parameters (such as temperature), thereby affecting its precision [16]. An appropriate selection of the optical system, together with a simplified demodulation method and the capability to suppress crosstalk, constitutes a critical factor in optical fiber sensing measurements.

In this work, polarization maintaining fiber (PMF) is used to connect to the Sagnac loop, and periodic troughs are used as monitoring signals. Optical fiber transmission offers the advantage of low loss in the near-infrared spectrum. Therefore, the wavelength range of 1400 nm to 1600 nm was selected as the monitoring band. The surface of PMF cladding was coated with a composite polymer film containing PVA and carbon nanosheets (CNs) as humidity-sensitive materials. Composite polymer films were deposited on the surface of PMF cladding using spin coating heating and laser processing in a high-temperature environment. The incorporation of composite polymers enables the film to exhibit multiple water absorption properties, while the surface functionalization through high temperature and laser heating imparts a fixed structure with micropores [17-18]. These enhancements can significantly enhance sensitivity, reduce signal-to-noise ratio, and contribute to precise humidity measurement. To achieve pH responsiveness, PAH and PAA were alternately deposited onto the PMF surface through electrostatic layer-by-layer assembly, resulting in a pH-sensitive nanocoating characterized by an undulating, island-like morphology that increases the effective contact area with the surrounding environment. Due to the inherent trough characteristics of PMF, there is no need for flat thin films as waveguide structures, which also provides conditions for the surface functionalization of humidity sensitive thin films. In order to overcome the crosstalk issue caused by temperature cross sensitivity, a temperature compensated PMF was connected in the same Sagnac loop. Based on the Vernier effect, different troughs exhibit different sensitivity characteristics to the same environmental parameter. Combined with the multi wavelength matrix, precise synchronous monitoring of humidity, pH and temperature can be achieved. Through a comprehensive comparison with related works from the past few years (Supplementary Table 1) [19-24], our study demonstrates that, while ensuring measurement accuracy and sensitivity, the system greatly simplifies complexity and reduces demodulation difficulty, thereby enabling real-time monitoring and achieving scalable measurement potential.

## 2. Experiment
### 2.1 Materials
Polyvinyl alcohol (PVA, Mw ~ 27,000, ≥99%), Poly(allylamine hydrochloride) (PAH, Mw ~ 15,000, ≥98%), Poly(acrylic acid) (PAA, Mw ~ 450,000, ≥99% Macklin) were purchased from MALKLIN. Carbon nanotubes (CNTs, NO: XFM40, >95% purity, Nanjing XFNANO Materials Tech Co., Ltd., China). Polarization maintaining fiber (Panda-type PMF, ChangFei Co., Ltd., China).

### 2.2 Preparation of composite polymer film
As shown in Fig. 1(a), 0.9 g PVA powder and 0.1 g carbon nanotubes were dissolved in 10 mL deionized water and magnetically stirred at room temperature for 24 h until the solute was completely dissolved. The mass fraction of the composite polymer was kept at 10%. The coating layer of PMF has been removed. The PMF was axially rotated and suspended while the composite polymer was encapsulated onto the cladding surface at 80℃ [25]. Then the PMF was heated in the temperature-controlled cabinet for 30 min at 120 ℃. The uniformly rotating PMF (30 r/min) was subjected to pulse laser ablation (532 nm Nd: YAG; Average power: 600 mW). Each spot was ablated for a duration of 10 min, and the composite polymer film was subsequently prepared on the cladding surface. By comparing Fig. 1(b) and Fig. 1(c), it can be clearly seen that the surface functionalized PVA/CNs composite polymer film has a surface microporous structure, which is due to the high energy acting on the polymer surface, resulting in a micro-melting effect. This will be very beneficial for improving sensing performance. Fig. 1c shows the similarity of the film morphologies prepared in two separate batches, demonstrating the reproducibility of the method. The PMF after surface functionalization is shown in the Fig. 1(d). The PMF has a diameter of 125μm, so the polymer film thickness ranges from 230μm to 270μm. As shown in Fig. 1(e), the PMF fiber was first immersed in the PAH solution (20 mM), allowing positively charged PAH molecules to adsorb onto the fiber surface. Subsequently, the fiber was rinsed with deionized water to remove any unbound molecules, ensuring the uniformity and integrity of the coating. The fiber was then immersed in the PAA solution (20 mM), where negatively charged PAA molecules electrostatically adsorbed onto the pre-deposited PAH layer, forming a stable bilayer structure. Afterward, a second rinsing step with deionized water was conducted to eliminate excess PAA molecules [26]. Fig. 1(f) and Fig. 1(g) show that repeat this step 50 times to fabricate a pH-sensitive polymer film with a thickness of 180 nm and an undulating, island-like surface structure, which facilitates enhanced contact area with the surrounding environment.

## 2.3 Measurement procedure

The broadband light source (BBS) was used as the input. The spectrometer (AQ6317B, wavelength resolution of 0.05 nm) was used to receive the signal. The highest resolution was 0.01 nm. It can fully display the spectral characteristics. The PMF (6cm) was connected to the Sagnac loop as shown in Fig 2(a). The optical fiber sensor was fixed onto fiber holders. Temperature control box (TCB) was used to realize the change of temperature. The humidity detector (HuaYi, PM6508) and the PMF were placed in the same position, and the humidifier was used to change the humidity of the environment. The pH value of the environment can be changed by adjusting the pH value of the solution inside the humidifier. When measuring pH sensitivity, keep the humidity constant. The pH tester (LiChen, pH-100B) was used to monitor the change of solution pH. Take the average value of multiple measurements as the recorded data.

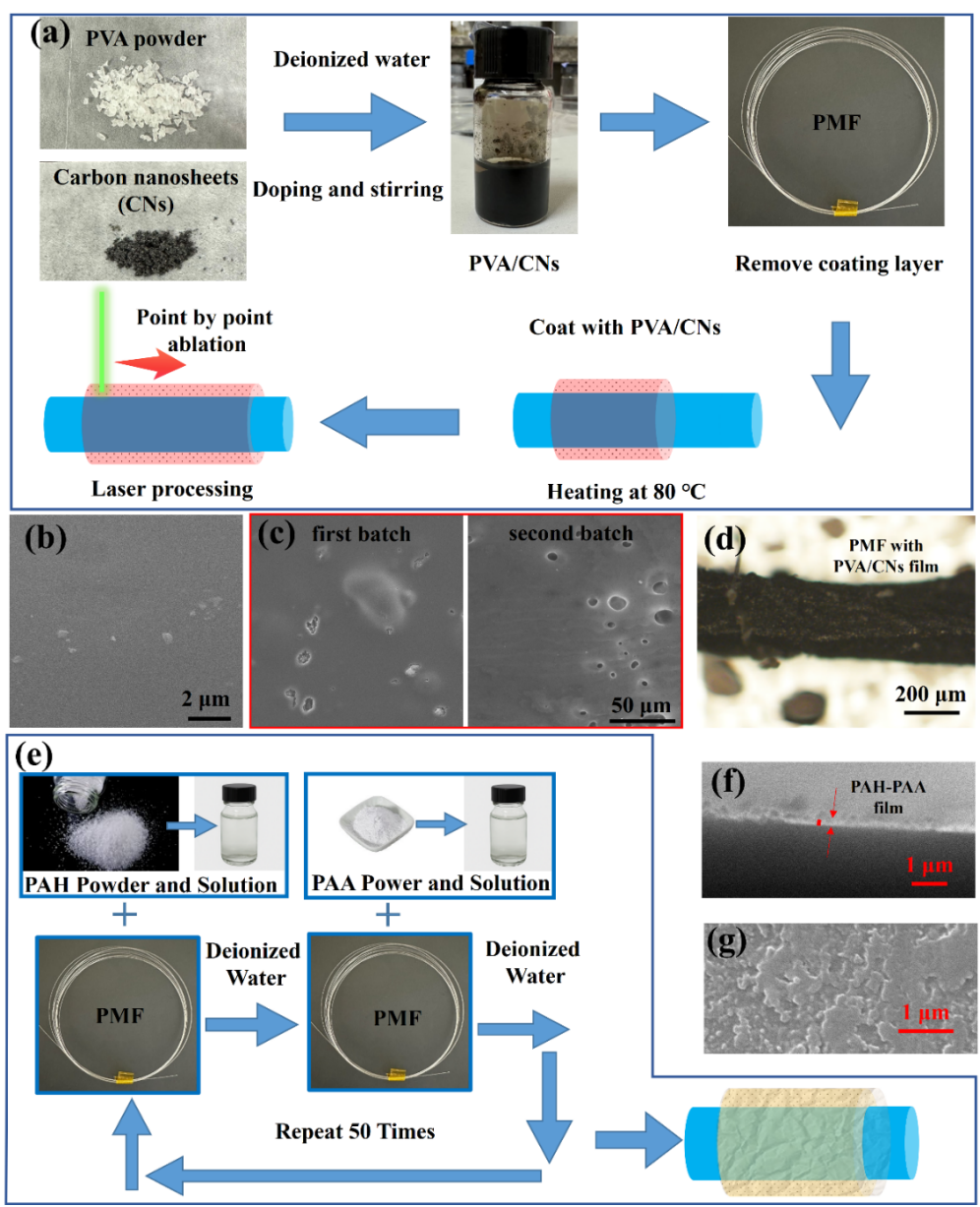

Fig. 1. (a) Functionalization process of PMF coated with a PVA/CNs film. (b) SEM

image of the composite polymer film without high-temperature heating and laser processing. (c) SEM image of the composite polymer film after high-temperature heating and laser processing. (d) Microscopic image of PMF coated with a PVA/CNs film. (e) Functionalization process of PMF coated with a PAH/PAA film. (f) Cross-sectional SEM image showing the thickness of the PAH/PAA film. (g) Surface morphology of the PAH/PAA film.

## 3. Results and discussion
### 3.1 Principle of humidity sensing
The polymer-coated PMF is connected to the Sagnac optical path in Fig. 2(a), and the incident light through the 3dB coupler will pass through the PMF in both clockwise and counterclockwise directions. The corresponding experimental physical diagram is shown in Fig. 2b. No polarization-maintaining (PM) components are required in any optical path, as shown by formula 1 [27],

$$T(\lambda) = \left[\sin(\theta)\cos\left(\frac{\pi BL}{\lambda}\right)\right]^2 \quad (1)$$

where L, B, and λ are the length, the birefringence, and the operation wavelength of the PMF, respectively. θ is the angle between the input light and the fast axis of the PMF. According to formula 2

$$\Delta\lambda \approx \frac{\lambda^2}{BL} \quad (2)$$

Δλ denotes the wavelength spacing between adjacent interference troughs. Due to the birefringence of PMF, the light propagating opposite introduces a relative phase difference, and the spectrometer receives a periodically modulated trough. The troughs are sensitive to temperature, stress and refractive index (RI), and the shift of trough can be used as a monitoring signal to react in real time. As shown in Fig. 2(c) and 2(d), when the ambient humidity increases, the PVA/CNs polymer will have a tendency to expand [28], when the ambient pH increases, the PAH-PAA polymer will have a tendency to shrink, which result in a change in the RI of the film. By formula 3 [29],

$$\frac{\Delta\lambda}{\lambda} = \frac{\partial \Delta n}{\Delta n} = \frac{1}{\Delta n}\frac{\partial \Delta n}{\partial RI}\Delta RI \quad （3）$$

Δn is the difference between the effective RI of the PMF and the RI of the external environment, so when the RI of the external environment changes, the trough of the PMF will shift accordingly. The effects caused by stress are discussed in detail in the Supplementary Information.

The humidity sensitivity (S) of the PMF sensor was determined by: S = Δλ / ΔRH, where Δλ is the wavelength shift and ΔRH is the change in relative humidity. The detection limit was evaluated using 3σ/S, where σ (0.02 nm) is the standard deviation of the wavelength resolution of spectrometer, and S is the humidity sensitivity. Accordingly, the sensitivity and resolution for temperature and pH are also calculated.

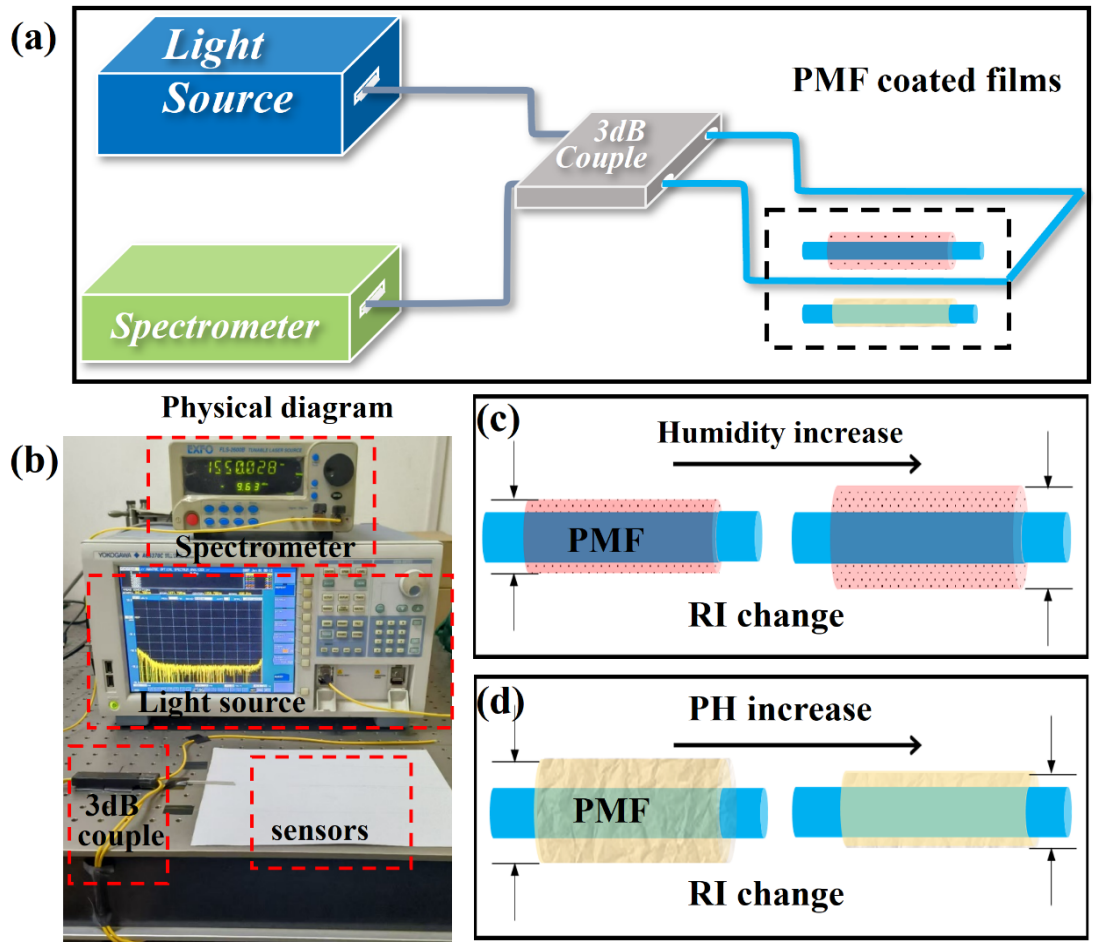

Fig. 2. (a) The PMF connection in Sagnac optical path diagram and composite humidity-sensitive polymer coated on the surface of PMF cladding. (b) Experimental physical diagram of PMF sensor system. (c) Schematic diagram of humidity-sensitive films changing with ambient humidity. (d) Schematic diagram of pH-sensitive films changing with ambient pH.

**3.2 Humidity sensitivity enhancement and temperature crosstalk compensation**

Fig. 3(a) shows the humidity sensing effect of PMF coated with composite polymer. One of the troughs is selected for monitoring, and the trough will shift towards the long wave direction with the increase of humidity, which reflects the humidity sensitivity of the polymer-coated PMF sensor. However, the traditional optical fiber humidity sensor is often affected by low sensitivity and multi-parameter crosstalk.

In order to improve the sensitivity of the sensor and reduce the signal-to-noise ratio, PVA doped CNs composite polymer was used to replace the traditional single humidity-sensitive polymer as the sensing film. Compared with the humidity sensitivity of PMF coated with PVA (red line in Fig. 3(b)), the humidity sensitivity of PMF coated with composite polymer (orange line in Fig 3(b)) has a significant increase from 0.12nm/% to 0.24 nm/% when the humidity is measured in the range of 34.5% to 90.4%. CNs are humidity-sensitive materials with high specific surface area, which are embedded in the polymer to achieve the effect of multiple water absorption [30]. However, due to the uneven dispersion of CNs caused by gravity deposition during the drying process of composite polymers, the linearity and sensitivity are not optimal.

In order to further increase the sensitivity and improve the linearity to increase the measurement accuracy, we optimized the preparation method of the composite polymer film. In order to achieve rapid formation of composite polymer films to reduce the deposition of CNs, the composite polymer is coated on the cladding surface at 80℃. After the laser ablation, the local overheating of the film causes the internal bubble to overflow, and microspores of the film in Fig 1(c) are formed, which increase the contact area between the internal film and the water molecules in the air, and the internal structure is more stable. As shown in the blue line in Fig 3(b), the humidity sensitivity is further improved to 0.26 nm/%, with a linearity of 0.99, and a measurement resolution (minimum detectable humidity interval) of 0.2 %RH. When the humidity is maintained at 49%, as shown in Fig. 3(c), the relative movement of the trough is only 0.3nm within 10 min, and the humidity accuracy can be maintained at 1.15%. The repeatability measurement results are shown in Fig. 3(d), and the sensor maintains the same sensitivity when the humidity is increased from 36.7% to 90.4% and decreased from 90.4% to 37.2%. The linearity is almost the same, and humidity sensing has good repeatability. After six months of storage, the pH sensor still retained 95% of its initial response (see Supplementary Figure 1).

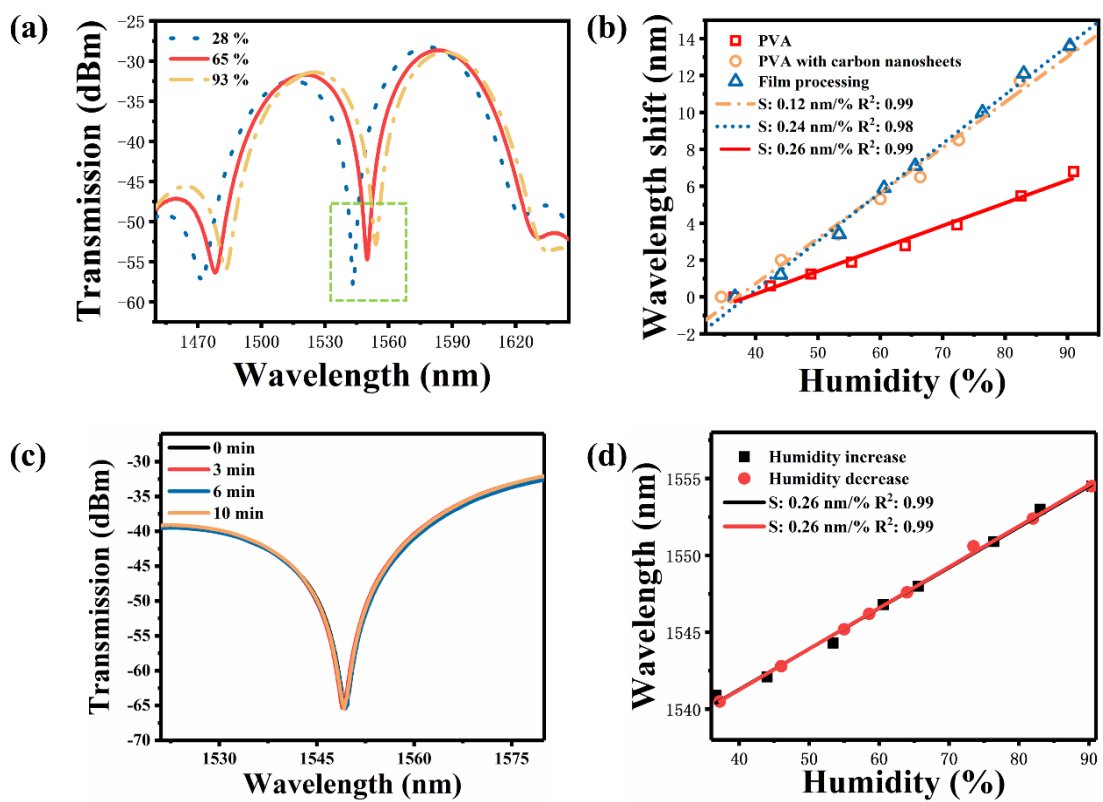

Fig. 3. (a) Spectral variation with humidity. (b) Linear comparison of sensitivity of PVA humidity-sensitive film (red), composite polymer humidity-sensitive film (yellow) and surface functionalized composite humidity-sensitive film (blue). (c) Time stability of surface functionalized PMF sensor. (d) Repeatable measurement of humidity increase and decrease.

In order to further improve the measurement accuracy of the sensor, it is necessary to eliminate the cross-stalk phenomenon of other parameters in the measurement process, especially the interference brought by the temperature change in the external environment to the humidity measurement. As shown in formula 4 [31],

$$\frac{d\lambda_T}{dT} = \frac{dB(T)}{dT}\frac{\lambda}{B(T)} + \frac{dL}{dT}\frac{\lambda}{L} \quad (4)$$

The change of ambient temperature will also cause the shift of PMF trough, and this is the intrinsic characteristic of PMF.

In order to simplify the complexity of multiple sensor multiplexing in previous work [32], cascaded PMF in the same Sagnac loop was used in this work to solve the temperature crosstalk problem. First, we illustrate the effectiveness of overcoming temperature crosstalk using the dual-wavelength matrix as an example. As shown in Fig. 4a, the PMF a is coated with the PVA/CNs composite polymer, and the PMF b is without any functionalization. The PMFs in the Sagnac loop were connected with a relative angle of approximately 45° between their fast axes to enhance polarization mode coupling and spectral sensitivity. The other parameters of the two PMF sensors are the same Therefore, both PMF sensors are sensitive to temperature, while only PMF a is sensitive to humidity. When the ambient temperature changes, both PMF sensors will play a role in the modulation of the spectrum, and when the ambient humidity changes, only PMF a plays a modulation role, and PMF b is used as a reference [33]. As shown the trough spacing (Δλ) in formula 2, each trough is modulated by two sensors. In cascaded PMF system, the different length (L) and birefringence (B) of the two sensors, result in slightly mismatched free spectral ranges between adjacent interferometers. This mismatch creates Vernier effect amplification, which enhances the relative sensitivity differences among troughs. [34-36]. When any environmental variable acts on the PMF alone, the sensitivity of each trough will be different, which provides the basis for implementing multi-wavelength matrices.

As shown in Fig. 4(b) and (d), two troughs for monitoring were selected. When the temperature increases, both troughs move in the short-wave direction, while when the ambient humidity increases, both troughs move in the long-wave direction. The variation in spectral shape exhibits slight differences when there are changes in temperature and humidity, which also demonstrates that the trough modulated by the PMF a and PMF b together is different from the trough change adjusted by the PMF b alone.

As depicted in Fig. 4(c) and 5(e), the temperature sensitivities of trough 1 and trough 2 in the cascade PMF system are determined to be -1.41 nm/℃ and -1.33 nm/℃ respectively, and the humidity sensitivities of trough 1 and trough 2 are found to be 0.29 nm/% and 0.23 nm/% respectively, exhibiting excellent linearity characteristics. Subsequently, the dual-wavelength matrix is substituted into formula 5 [27],

$$\begin{bmatrix} \Delta\lambda_{Trough\ 1} \\ \Delta\lambda_{Trough\ 2} \end{bmatrix} = \begin{bmatrix} K_{T_{Trough\ 1}} & K_{H_{Trough\ 1}} \\ K_{T_{Trough\ 2}} & K_{H_{Trough\ 2}} \end{bmatrix} \begin{bmatrix} \Delta T \\ \Delta RH \end{bmatrix} \quad (5)$$

$K_{T_{Trough\ 1}}$ is the sensitivity of trough 1 to change with temperature, $K_{H_{Trough\ 1}}$ is the sensitivity of trough 1 to change with humidity, and $K_{T_{Trough\ 2}}$ is the sensitivity of trough 2 to change with temperature. $K_{H_{Trough\ 2}}$ is the sensitivity of trough 2 to change with humidity. The following formula 6 is obtained.

$$\begin{bmatrix} \Delta\lambda_{Trough\ 1} \\ \Delta\lambda_{Trough\ 2} \end{bmatrix} = \begin{bmatrix} -1.41 & 0.29 \\ -1.33 & 0.23 \end{bmatrix} \begin{bmatrix} \Delta T \\ \Delta RH \end{bmatrix} \quad (6)$$

The negative sign indicates the movement of the trough towards the short-wave direction, while a positive sign signifies its movement towards the long wave direction. By observing changes in these two troughs, simultaneous monitoring of temperature and humidity can be achieved using the cascaded PMF sensor without requiring additional sensors, ultimately enabling high-precision real-time humidity monitoring. The coefficients of each trough with respect to temperature, humidity, and pH are different, and these differential responses ensure the specificity and selectivity of the measurements, as the multi-wavelength matrix can uniquely decouple the contributions of each parameter. [37-38]

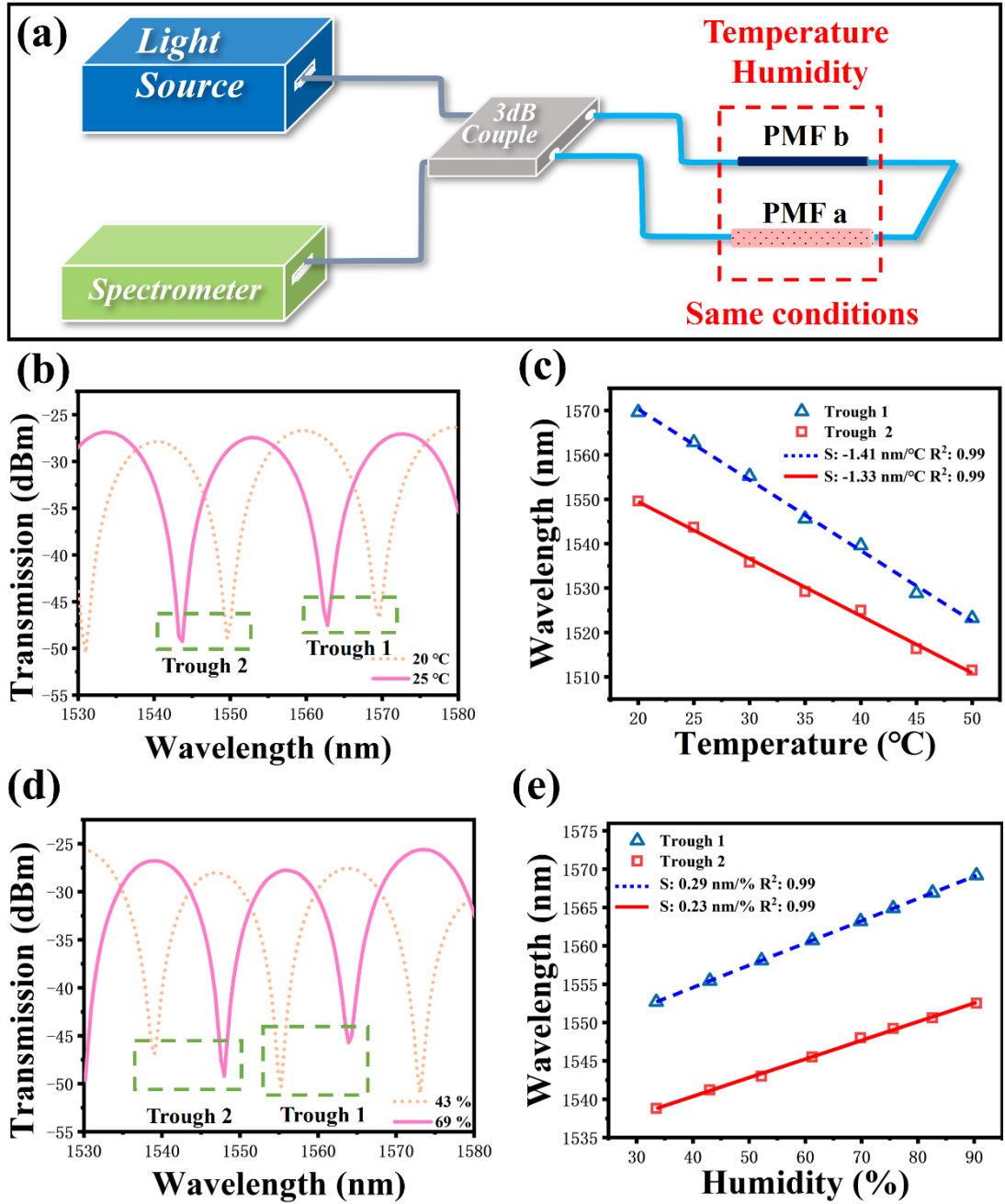

Fig. 4. (a) PMF a and PMF b sensor cascade system (b) Cascaded PMF spectral variation with temperature. (c) Dual trough linear temperature sensitivity relationship. (d) Cascaded PMF spectral variation with humidity. (e) Dual trough linear humidity sensitivity relationship.

### 3.3 Humidity, pH and temperature synchronous measurement system

Fig. 5(a) illustrates the spectral response of the PMF sensor under varying pH environments. As the pH value increases, the PAH-PAA film undergoes shrinkage, resulting in a change in the composite RI. This leads to a redshift of the trough in the transmission spectrum, demonstrating the excellent pH sensitivity. Figure 5(b) presents the linear fitting results between the troughs and pH values, based on two representative troughs. A clear linear trend is observed with increasing pH, exhibiting sensitivity values of approximately 1.02 nm/pH and 1.03 nm/pH, respectively, with a measurement resolution of 0.05 pH. The linearity with R² = 0.99 confirms that the sensor possesses an excellent linear response. As shown in Figure 5(c), the sensor's temporal stability was evaluated at a fixed pH of 4.3. Over a continuous monitoring period of 10 minutes, the spectral dip exhibited minimal drift—only about 0.25 nm—indicating outstanding time stability. Figure 5(d) demonstrates the reproducibility test results under cyclic pH changes. The pH value was increased from 3.6 to 7.2 and then decreased back to 3.6. The troughs shift during the pH increase and decrease processes, followed nearly identical trends, and the sensitivity remained unchanged. After six months of storage, the pH sensor still retained 95% of its initial response (see Supplementary Figure 2). This confirms that the PMF pH sensor offers excellent repeatability, making it suitable for practical applications involving multiple usage cycles.

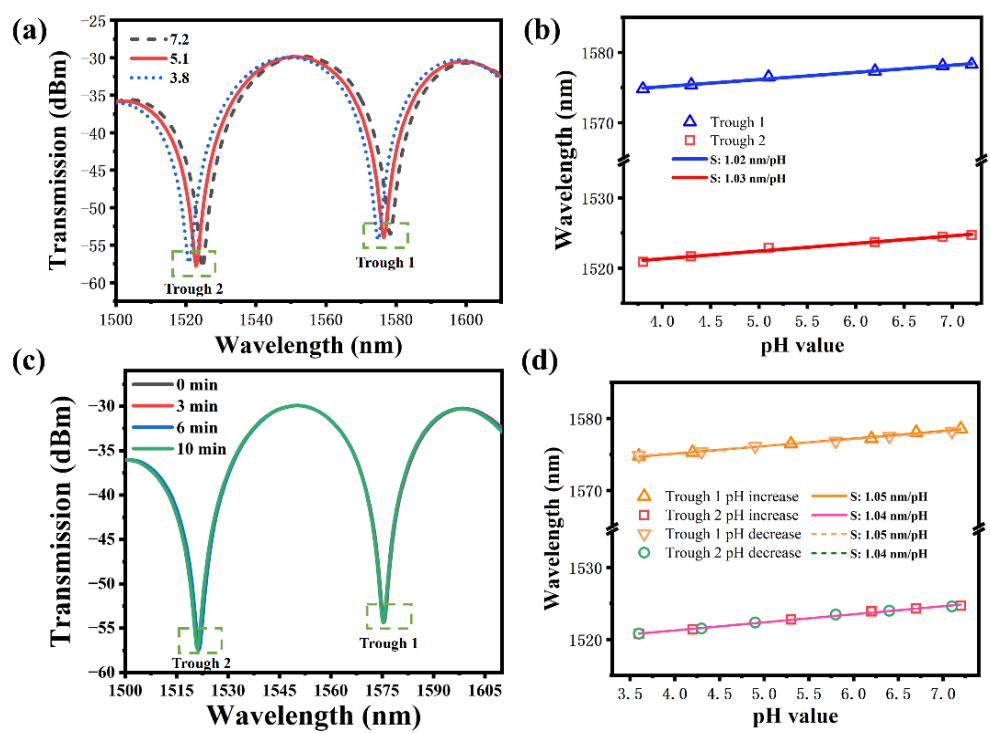

fig. 5. (a) Spectral variation with pH. (b) Linear comparison of sensitivity of PAH-PAA pH-sensitive film (c) Time stability of PAH-PAA functionalized film PMF sensor. (d) Repeatable measurement of pH increase and decrease.

Figure 6 (a) shows a schematic diagram of the structure of a multi-parameter synchronous monitoring system. Follow the same method as in section 3.2, as shown in Fig. 6a, the PMF a is coated with the PVA/CNs composite polymer, and the PMF b is without any functionalization. PMF c is coated with the PAH-PAA nano films. The three are connected in series to form a cascade structure, which achieves synchronous detection of humidity, pH, and temperature through a multi-wavelength matrix. Three troughs for monitoring were selected. Measure the response of three troughs to variations in temperature, humidity, and pH using the method described in Section 3.2 (results in the Fig. 6 b-g), and incorporate the results into Formula 7 for the three-wavelength matrix calculation. It is worth noting that the pH sensitivity of the two troughs in Fig. 5b is almost the same, while there is a significant difference in the pH sensitivity of the three troughs after cascading (Fig. 6f and g), which proves the role of the Vernier effect under this cascading structure. The specific demodulation mechanism of the multi-wavelength matrix can be found in the Supporting Information.

$$\begin{bmatrix} \Delta\lambda_{Trough\ 1} \\ \Delta\lambda_{Trough\ 2} \\ \Delta\lambda_{Trough\ 3} \end{bmatrix} = \begin{bmatrix} -1.59 & 0.31 & 1.03 \\ -1.37 & 0.23 & 1.12 \\ -1.45 & 0.17 & 0.94 \end{bmatrix} \begin{bmatrix} \Delta T \\ \Delta RH \\ \Delta pH \end{bmatrix} \quad (7)$$

In terms of measurement performance, the system demonstrates good sensitivity (1.45 nm/°C, 0.23 nm/%RH, and 1.03 nm/pH) and high accuracy (0.04 °C, 0.26 %RH, and 0.05 pH), with excellent repeatability and reusability (maintained for at least six months). A detailed explanation of the measurement range is provided in the Supplementary Information. The cascaded structure of functionalized PMF combined with multi-wavelength matrices has been verified to have excellent universal applicability and can be used to solve various crosstalk problems in fiber optic sensing. Notably, the measurement method is scalable to higher-order matrix forms by including additional spectral dips, enabling multi-parameter sensing beyond humidity, pH, and temperature. However, it should also be noted that our current experiments were conducted under controlled laboratory conditions. In more complex real-world environments (such as industrial monitoring), this cascaded PMF structure inherently provides resistance to electromagnetic interference and structural stability due to its all-fiber design. Future work will focus on long-term field validation.

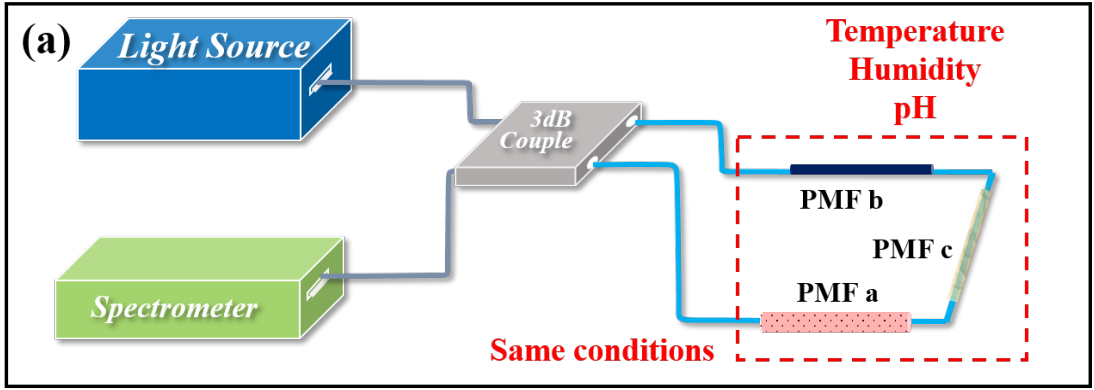

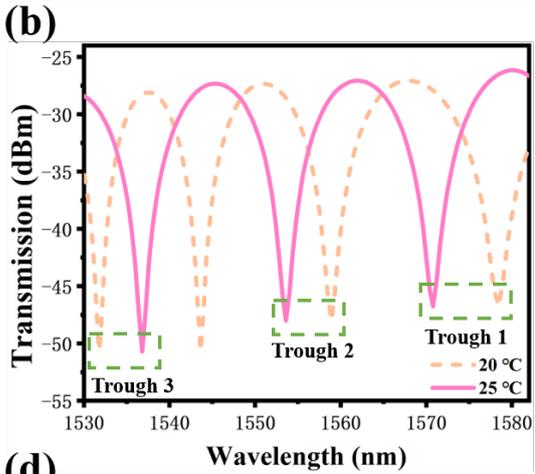
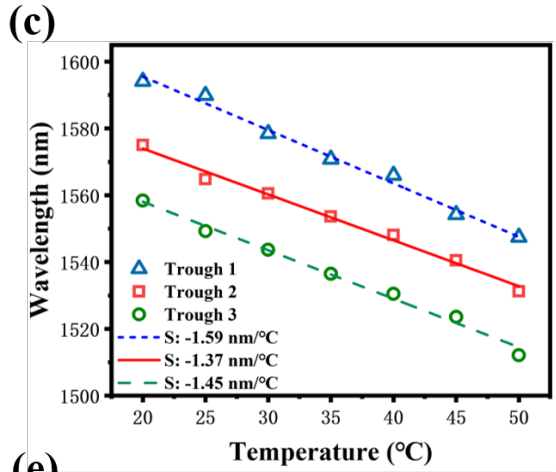

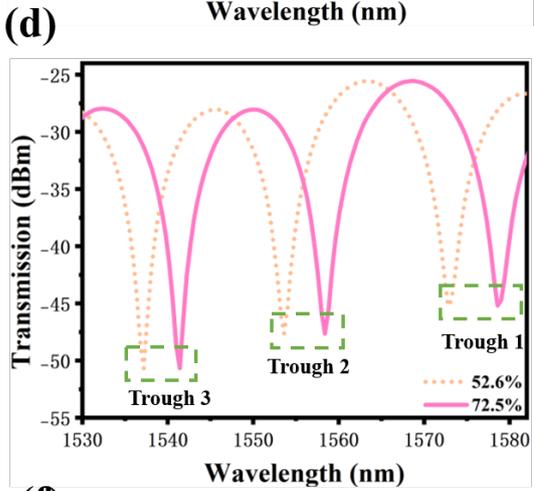
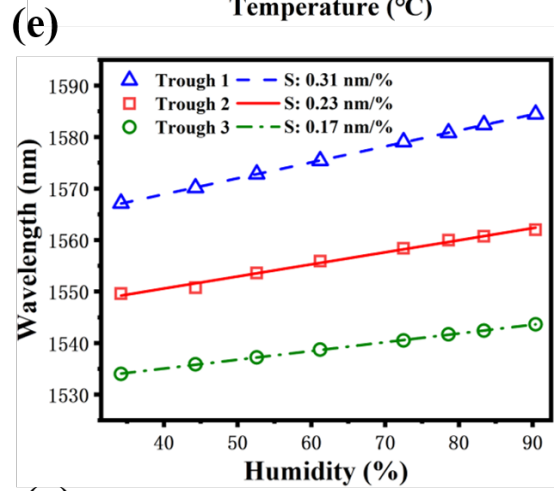

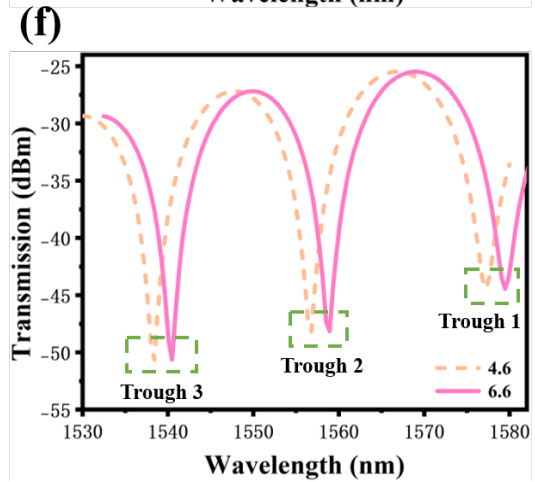
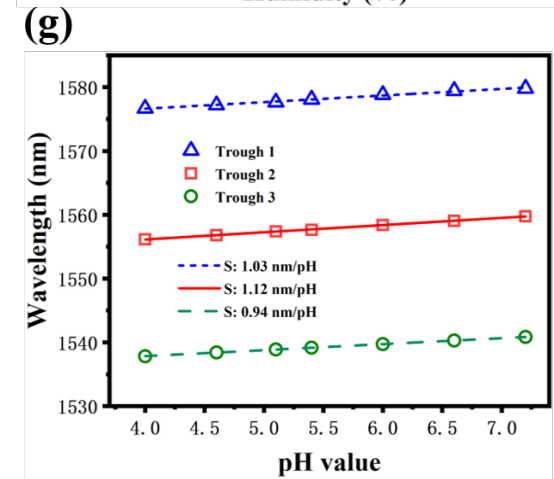

Fig. 6. (a) PMF a, PMF b and PMF c sensor cascade system (b) Cascaded PMF spectral variation with temperature. (c) Three troughs linear temperature sensitivity relationship. (d) Cascaded PMF spectral variation with humidity. (e) Three troughs linear humidity sensitivity relationship. (f) Cascaded PMF spectral variation with pH. (g) Three troughs linear pH sensitivity relationship.

## 4. Conclusion

In this work, we used PMF as the carrier, the composite polymer composed of PVA and CNs was functionalized on its cladding surface, and the improved method of high temperature rapid film formation and laser processing was used to prepare a micro-porous humidity sensitive film. The humidity sensitivity of the sensor increased from 0.12 nm/%RH to 0.26 nm/%RH, with improved stability and a measurement resolution of 0.2% relative humidity. PH-sensitive thin films were prepared using a fast and efficient layer-by-layer method. The pH sensitivity of the sensor was about 1.03 nm/pH, with a measurement resolution of 0.05 pH, within the pH range of 3.8-7.2. The cascaded PMF sensor system is further used to monitor the three troughs. By relying on the vernier effect and combining the signal processing method of multi-wavelength matrix, the precise measurement of temperature, pH and humidity was achieved, the problem of temperature crosstalk was solved, and finally the multi-parameter synchronous monitoring system was realized.

**Acknowledgements**
This work was supported by the Shandong Agriculture and Engineering University Start-Up Fund for Talented Scholars（BSQJ202309）